\documentclass[aps,showpacs,superscriptaddress,twocolumn]{revtex4}%
\usepackage{amsfonts}
\usepackage{amsmath}
\usepackage{amssymb}
\usepackage{graphicx}%
\begin{document}
\title{Exciton condensation and fractional charge in a bilayer two-dimension electron
gas adjacent to a superconductor film}
\author{Ningning Hao}
\affiliation{Institute of Physics, Beijing 100190, People's Republic of China}
\affiliation{Institute of Applied Physics and Computational Mathematics, P.O. Box 8009,
Beijing 100088, People's Republic of China}
\author{Ping Zhang}
\thanks{Corresponding author. zhang\_ping@iapcm.ac.cn}
\affiliation{Institute of Applied Physics and Computational Mathematics, P.O. Box 8009,
Beijing 100088, People's Republic of China}
\affiliation{Center for Applied Physics and Technology, Peking University, Beijing 100871,
People's Republic of China}
\author{Jian Li}
\affiliation{Institute of Physics, Beijing 100190, People's Republic
of China}
\author{Wei Zhang}
\affiliation{Institute of Applied Physics and Computational Mathematics, P.O. Box 8009,
Beijing 100088, People's Republic of China}
\author{Yupeng Wang}
\affiliation{Institute of Physics, Beijing 100190, People's Republic
of China}


\begin{abstract}
We study the exciton condensate (EC) in a bilayer two-dimension-electron-gas
(2DEG) adjacent to a type-II superconductor thin film with an array of pinned
vortex lattices. By applying continuum low energy theory and carrying
numerical simulations of lattice model within mean-field approximation, we
find that if the order parameter of EC has a vortex profile, there are exact
zero modes and associated \emph{rational} fractional charge for zero
pseudospin potential ($\mu$) and average chemical potential ($h$): $\mu$=$0$
and $h$=$0$; while for $\mu\mathtt{\neq}0$ and $h$=$0$, intervalley mixing
splits the zero energy levels, and the system exhibits \emph{irrational}
fractional \emph{axial} charge.

\end{abstract}
\maketitle

\textbf{Introduction} Charge fractionalization and fractional statistics
\cite{wilczek} in two-dimensional systems have attracted much attention due to
their potential applications in topological quantum computation
\cite{fw1,fw2,fw3}. One representative example is the well known $\nu$%
=$\frac{5}{2}$ fractional quantum Hall $($FQH$)$ system \cite{Read}. The
excitations of the $\nu$=$\frac{5}{2}$ FQH system carry fractional charge and
exhibit non-Abelian fractional statistics. The pivotal features of FQH system
are strong correlations and broken time-reversal symmetry $($TRS$)$ as a
result of the electrons' strong coulomb interaction and strong external
magnetic field. However, some recently proposed models \cite{Hou,Franz} have
revealed that the strong correlations and broken TRS are not necessary for
fractionalization. Moreover, these models indicate that the nontrivial
topological configurations play a key role in the fractionalization. The
common features of these models are that the TRS is not broken, and the low
energy excitations can be described by the Dirac field coupling with a
topologically nontrivial scalar field (vortex profiles, for instance)
\cite{Jackiw}. Therefore, the energy spectrum's symmetry and index theorem of
the Dirac operator protect the zero modes and the associated fractionalization
\cite{Chamon}. In another interesting model \cite{Weeks} different from
aforementioned ones, the fractionalization of the excitations is due to the
response to the half-magnetic-flux-quanta defect of the system in the integer
quantum hall (IQH) regime. This model is more likely to be fabricated compared
to aforementioned systems.

We start with a bilayer two-dimension-electron-gas (2DEG), as illustrated
schematically in Fig. 1(a). Each layer is adjacent to a film of type-$\Pi$
superconductor supplying the quantization of flux in unit of $\frac{1}{2}%
\Phi_{0}$ ($\Phi_{0}$=$h/e$) by the Abrikosov vortex square lattices. The
spin-polarized electrons in the bilayer are almost localized near the
Abrikosov vortex cores. At half filling (namely, the number of electrons is
one half of the total lattice sites), the bilayer system has a magnetic
filling factor of two, and the magnetic flux through one square lattice cell
is $\pi$. In addition, the two layers are separated by a spacer which is
usually a dielectric barrier (e.g., SiO$_{2}$). The chemical potential in each
layer can be adjusted independently by the respective gate voltage. The
carrier in each layer has electron or hole nature determined by the respective
band structure. In particular, the inverse external gate voltages in two
layers bias the charge balance of two layers and result in the perfect
particle-hole symmetry. So the electrons in one layer and holes in another
layer bound together and form exciton condensate $($EC$)$ due to the
interlayer coulomb interaction. Our system is favorable for EC because of the
strong magnetic field and low electron density as suggested by experiments of
the coupled quantum wells \cite{Fuku}\cite{Eisen}.

\begin{figure}[ptb]
\begin{center}
\includegraphics[width=1\linewidth]{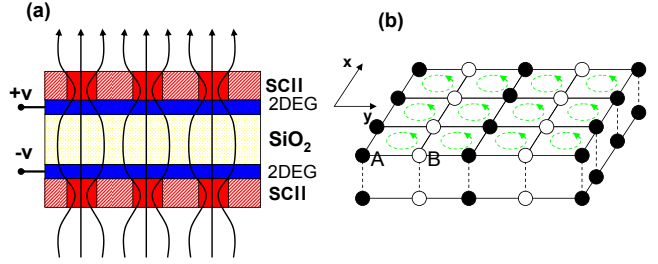}
\end{center}
\caption{(Color online) (a) Schematic structure of exciton condensate system.
The external gates induce the symmetric electron and hole carriers. (b) Our
bilayer square lattice model. }%
\end{figure}

In this paper, we construct a tight-binding model analogous to bilayer
graphene \cite{Franz1} and solve the Bogoliubov-de Gennes (BdG) equation
self-consistently in real space to determine the amplitude of EC order
parameter (OP) and study the system's topological properties originating from
OP with vortex configuration. The low energy excitations of the system present
novel features due to the vortex configuration in OP. We show that when the
biased voltage is zero, two degenerate fermion zero modes (each one for one
valley) emerge. Furthermore, for the vortex with vorticity ($n$=1) at half
filling, each layer has fractional charge $\frac{1}{2}e$ or $-\frac{1}{2}e$
(depending on whether the two zero modes are occupied or not), though the
total and difference charge of bilayer are the integer numbers. When the
biased voltage is not zero, the two fermion zero modes slightly and
symmetrically split around zero energy due to the intervalley mixing. At half
filling, the axial charge of the whole system is fractional just as what
happens in bilayer graphene studied in Ref. (13).

\textbf{Lattice model and exciton condensation} In the following, we present a
model to describe the EC in our system at near zero temperature. The essential
terms include intralayer hopping between the nearest neighbor sites. We ignore
the intralayer carriers' Coulomb interactions because their effects only
renormalize the energy bands. We consider interlayer Coulomb repulsion as an
effective short-range repulsion only between interlayer electrons at the same
planar sites. Furthermore, the external biased voltage is included. Since the
electron spin is polarized along the direction of the magnetic field, we omit
spin index. There is no direct hopping between the two layers.

The Hamiltonian of our system can be expressed as follows%
\begin{align}
&  H=\underset{i\alpha}{-\sum}\mu_{\alpha}n_{i\alpha}-\underset{<i,j>\alpha
}{\sum}t_{ij\alpha}e^{i\theta_{ij\alpha}}c_{j\alpha}^{+}c_{i\alpha}\nonumber\\
&  +U\underset{i}{\sum}n_{i1}n_{i2}.\label{wholeH}%
\end{align}
Here $c_{i\alpha}$ annihilates an electron at site $i$ of layer-$\alpha$
($\alpha$=$1,2$), $n_{i\alpha}$=$c_{i\alpha}^{+}c_{i\alpha}$, and
$t_{ij\alpha}$ is intralayer nearest-neighbor hopping with $t_{ij1}$=$t_{ij2}$
for the electron-electron bilayer and $t_{ij1}$=$-t_{ij2}$ for electron-hole
bilayer. We consider the electron-hole bilayer and set $t_{ij1}$%
=$-t_{ij2}\mathtt{\equiv}t$=$1$. $\mu_{\alpha}$ is the chemical potential of
layer-$\alpha$. The magnetic field is included through the Peierls phase
factors, $\theta_{ij\alpha}$=$\frac{2\pi}{\Phi_{0}}\int_{\mathbf{r}_{i\alpha}%
}^{\mathbf{r}_{j\alpha}}\mathbf{A}_{\alpha}\mathtt{\cdot}d\mathbf{l}$. We
choose to work in Landau gauge $\mathbf{A}_{\alpha}$=$\frac{1}{2}\Phi
_{0}(-y,0),$ where we have set the lattice space unity. The mean-field order
parameter of the EC is $\Delta_{i}$=$U\left\langle c_{i2}^{+}c_{i1}%
\right\rangle $ , and the mean-field Hamiltonian is%
\begin{align}
H_{MF} &  =-\underset{i\alpha}{\sum}\mu_{\alpha}n_{i\alpha}+\underset
{<i,j>\alpha}{\sum}(-)^{\alpha}te^{i\theta_{ij\alpha\beta}}c_{j\alpha}%
^{+}c_{i\alpha}\nonumber\\
&  -\underset{i}{\sum}(\Delta_{i}c_{i1}^{+}c_{i2}+h.c.)+\frac{1}{U}%
\underset{i}{\sum}\left\vert \Delta_{i}\right\vert ^{2}.\label{mfH1}%
\end{align}
With the help of the following Bogoliubov transformation%
\begin{equation}
\left[
\begin{array}
[c]{c}%
c_{i1}\\
c_{i2}%
\end{array}
\right]  =\underset{n}{\sum}\left[
\begin{array}
[c]{cc}%
u_{i}^{n} & -\upsilon_{i}^{n\ast}\\
\upsilon_{i}^{n} & u_{i}^{n\ast}%
\end{array}
\right]  \left[
\begin{array}
[c]{c}%
\alpha_{n}\\
\beta_{n}^{+}%
\end{array}
\right]  ,\label{Bogv}%
\end{equation}
the above Hamiltonian(\ref{mfH1}) can be diagonalized by solving the following
BdG equation%
\begin{equation}
\underset{j}{\sum}\left[
\begin{array}
[c]{cc}%
h_{ij,1} & -\Delta_{i}\delta_{ij}\\
-\Delta_{i}^{\ast}\delta_{ij} & h_{ij,2}%
\end{array}
\right]  =E_{n}\left[
\begin{array}
[c]{c}%
u_{j}^{n}\\
\upsilon_{j}^{n}%
\end{array}
\right]  ,\label{BdG}%
\end{equation}
where $h_{ij,1(2)}=\mp t-\mu_{1(2)}\delta_{ij}$. The self-consistent equation
of the OP is%
\begin{equation}
\Delta_{i}=-\underset{n}{\sum}u_{i}^{n}\upsilon_{i}^{n\ast}\tanh(\frac{E_{n}%
}{2k_{B}T}).\label{op}%
\end{equation}

We solve the set of BdG equations self-consistently via exact diagonalization
in real space. The system size of $24\mathtt{\times}24$ is used in the
calculation and the convergence criterion of $\Delta_{i}$ is set to 10$^{-4}$
in unit of the intralayer nearest-neighbor hopping. We consider the near-zero
temperature and set $k_{B}T$ to be 10$^{-3}$. For convenience of discussion,
we define the average chemical potential and pseudospin polarization potential
as $h$=$(\mu_{1}\mathtt{+}\mu_{2})/2$ and $\mu$=$(\mu_{1}\mathtt{-}\mu_{2}%
)/2$. We find that if $h$=$0$ (we set $\mu_{2}$=$-\mu_{1}$=$-\mu$), the
exciton order parameter is uniform: $\Delta_{i}$=$\Delta$=$m$, where $m$ is
real. Our calculations show that as $h$ decreases from zero, the phases of
system undergo BCS-like phase, 1D modulated Larkin-Ovchinnikov-like (LO) phase
\cite{LO}, 2D modulated LO-like phase and finally OP zero.

In the following, we restrict ourselves to the case of $\mu_{2}$=$-\mu_{1}%
$=$-\mu$. In the Landau gauge that we choose, the system can be viewed as
simplified lattices as depicted in Fig. 1(b). The magnetic unit cell in layer
$\alpha$ includes two sites $(l,n)\alpha$ and $(l,n$+$1)\alpha$ denoted by
$A\alpha$ and $B\alpha$. The wave functions are of the form of spinor field
$\psi(\mathbf{r})$=$[c_{B1}(\mathbf{r}_{l,n+1}),c_{A1}(\mathbf{r}%
_{l,n}),c_{B2}(\mathbf{r}_{l,n+1}),c_{A2}(\mathbf{r}_{l,n})]^{T}$. In the
momentum space with the reduced Brillouin zone $BZ$=$\{k_{y}\mathtt{-}%
\pi/2\mathtt{\leq}k_{x}\mathtt{\leq}k_{y}$+$\pi/2,\left\vert k_{y}\right\vert
\mathtt{\leq}\pi\}$, the Hamiltonian can be written as $H_{MF}$=$\sum
_{\mathbf{k}}\psi^{+}(\mathbf{k})\mathcal{H}_{MF}(\mathbf{k})\psi(\mathbf{k}%
)$+$E_{0}$ with $E_{0}$= $N_{0}m^{2}/U$ and
\begin{align}
\mathcal{H}_{MF}(\mathbf{k}) &  =-\mu\sigma_{3}\otimes I\mathbb{+}2\cos
(k_{x})\sigma_{3}\otimes s_{3}-2\cos(k_{y})\sigma_{3}\otimes s_{1}\nonumber\\
&  -me^{-i\chi\sigma_{3}\otimes I}\sigma_{1}\otimes I,\label{mfH2}%
\end{align}
where $\sigma_{i}$ $(s_{i})$ $(i$=$1,2,3)$ is Pauli matrix acting in
sublattice (pseudospin) space and $I$ is unitary matrix. Without loss of
generality, we preserve the constant phase $\chi$ of OP. This Hamiltonian
possesses the particle-hole symmetry $\ \ \ \ \ $%
\begin{equation}
\Omega\mathcal{H}_{MF}(\mathbf{k})\Omega=-\mathcal{H}_{MF}^{\ast}%
(-\mathbf{k})\label{sym1}%
\end{equation}
with $\Omega$=$-\sigma_{2}\mathtt{\otimes}I$ and $\Omega^{2}$=$1$. As a result
of this symmetry, it is easy to find%
\begin{equation}
\mathcal{H}_{MF}\Omega\psi^{\ast}=-\Omega\mathcal{H}_{MF}^{\ast}\psi^{\ast
}=-E\Omega\psi^{\ast}.\label{sym11}%
\end{equation}
It means that if $\psi$ is an eigenvector of $\mathcal{H}_{MF}$ with
eigenvalue $E,$ then $\Omega\psi^{\ast}$ is guaranteed to be an eigenvector of
$\mathcal{H}_{MF}$ with eigenvalue $-E$. So, all nonzero eigenvalues of
$\mathcal{H}_{MF}$ come in $\pm E$ pairs and the energy spectrum is symmetric
about zero energy. For uniform order parameter $m$, the energy spectrum is of
the form
\begin{equation}
E(\mathbf{k})=\pm2\sqrt{(\sqrt{\cos^{2}k_{x}+\cos^{2}k_{y}}\pm\mu)^{2}+m^{2}%
}.\label{energy}%
\end{equation}
The energy spectrum symmetry is clear in this case. Note that no in-gap
eigenvalues exist as $m\mathtt{\neq}0$ in uniform case.

\textbf{Zero modes and charge fractionalization }We now investigate the
low-energy behaviors of our system. In the low energy limit, the properties of
the system are dominated by the excitations around the two inequivalent
valleys $\mathbf{K}_{\pm}$=$(\mp\frac{\pi}{2},\mp\frac{\pi}{2})$ where
$\cos\left.  k_{x}\right\vert _{\mathbf{K}_{\pm}+\mathbf{p}}$=$\pm p_{x}$ and
$\cos\left.  k_{y}\right\vert _{\mathbf{K}_{\pm}+\mathbf{p}}$=$\pm p_{y}$.
Then, the linearized mean-field Hamiltonian for these excitations are
$\mathcal{H}_{MF+}$+$\mathcal{H}_{MF-}$ with $\mathcal{H}_{MF+}$%
=$-iI\mathtt{\otimes}\sigma_{2}\mathcal{H}_{MF-}iI\mathtt{\otimes}\sigma
_{2}\mathtt{\equiv}\mathcal{H}$,%
\begin{equation}
\mathcal{H}=-\mu\sigma_{3}\otimes I\mathbb{+}2p_{x}\sigma_{3}\otimes
s_{3}-2p_{y}\sigma_{3}\otimes s_{1}-me^{-i\chi\sigma_{3}\otimes I}\sigma
_{1}\otimes I. \label{conh}%
\end{equation}
Here, $\mathbf{p}$\textbf{=}$-i\nabla$ is the momentum operator. Since the
valleys are decoupled in the low-energy theory, we first study a single valley.

Around the valleys, the Hamiltonian $\mathcal{H}$ possesses a symmetric
property%
\begin{equation}
\Xi\mathcal{H}(\mathbf{K}_{\pm}+\mathbf{p})\Xi=\mathcal{H}^{\ast}%
(\mathbf{K}_{\pm}+\mathbf{p}), \label{sym2}%
\end{equation}
where $\Xi$=$i\sigma_{2}\mathtt{\otimes}s_{2}.$ This symmetry is unbroken only
around each valley. The low-energy excitations obey this symmetry around each
valley. The case of $\mu$=$0$ is of particular interest. In this situation,
the symmetry of our system gets enhanced. It is easy to check that
$\Theta\mathcal{H}\Theta$=$-\mathcal{H}$, $\Theta^{2}$=$1$, with
$\Theta\mathtt{\equiv}\sigma_{3}\mathtt{\otimes}s_{2}$. This symmetry has
important consequences on the number of zero modes.

Now suppose $\Delta$ is also taken to vary with position and consider OP with
a symmetric vortex of vorticity $n$, which means $\Delta(r,\theta
)$=$m(r)e^{in\theta}$ with $m(r)$ a real function of $r$ and the constant
phase $\chi$ omitted by the special gauge choice. In order to get explicitly
analytic solutions, we use the well defined vortex profile
\begin{equation}
\Delta(r,\theta)=\left\{
\begin{array}
[c]{c}%
m(r)e^{in\theta},r>0\\
0,r=0
\end{array}
\right.  , \label{profop}%
\end{equation}
In quantum limit, $m(r)$=$m_{0}$, where $m_{0}$ is a positive real constant.
The zero-mode solutions satisfy $\mathcal{H}\psi_{0}$=$0$. Eq. (\ref{sym2})
means $\mathcal{H}\Xi\psi^{\ast}$=$-\Xi\mathcal{H}^{\ast}\psi^{\ast}$%
=$-E\Xi\psi^{\ast}$. For zero modes, we impose $\Xi\psi_{0}^{\ast}$=$\psi
_{0}.$ If $\Xi\psi_{0}^{\ast}$=$-\psi_{0}$, then the transformation $\psi
_{0}\mathtt{\rightarrow}i\psi_{0}$ brings it back to $\Xi\psi_{0}^{\ast}%
$=$\psi_{0}$. So, in the zero-energy subspace, we can suppose $\psi_{0}%
$=$[f,g,ig^{\ast},-if^{\ast}]^{T}$. Hence, we can get two independent
equations for zero-mode solutions in quantum limit%
\begin{align}
(-\mu-2i\partial_{x})f+2i\partial_{y}g-im_{0}e^{in\theta}g^{\ast}  &
=0\nonumber\\
(-\mu+2i\partial_{x})g+2i\partial_{y}f+im_{0}e^{in\theta}f^{\ast}  &  =0.
\label{equs}%
\end{align}

In order to get the symmetric solutions in the presence of the symmetric
vortex profile, we introduce new wave functions $\zeta$ and $\xi$ defined by
the following equations,
\begin{equation}
\left[
\begin{array}
[c]{c}%
f\\
g\\
ig^{\ast}\\
-if^{\ast}%
\end{array}
\right]  =\frac{\sqrt{2}}{2}\left[
\begin{array}
[c]{c}%
-e^{-i\pi/4}(\zeta+\xi)\\
e^{i\pi/4}(\zeta-\xi)\\
e^{i\pi/4}(\zeta^{\ast}-\xi^{\ast})\\
-e^{-i\pi/4}(\zeta^{\ast}+\xi^{\ast})
\end{array}
\right]  . \label{transf}%
\end{equation}
Then, $\zeta,\xi$ satisfy the following equations in polar coordinate,%
\begin{align}
e^{i\theta}(\partial_{r}+ir^{-1}\partial_{\theta})\zeta-m_{0}/2e^{in\theta
}\zeta^{\ast}-i\mu/2\xi &  =0\nonumber\\
e^{-i\theta}(\partial_{r}-ir^{-1}\partial_{\theta})\xi+m_{0}/2e^{in\theta}%
\xi^{\ast}-i\mu/2\zeta &  =0, \label{equ2}%
\end{align}
which are almost the same as that in Ref. (13). The analogous solutions can be
gotten as%
\begin{equation}
\zeta=e^{-m_{0}r/2}J_{p}(\mu r/2)e^{ip\theta}\text{ and }\xi=-ie^{-m_{0}%
r/2}J_{q}(\mu r/2)e^{iq\theta}, \label{solu}%
\end{equation}
which are exponentially decayed Bessal functions and $p$=$q\mathtt{-}1$%
=$\frac{n-1}{2}$. So, the single valued solutions survive only when $n$ is
odd. Hence, one zero-mode solution $\psi_{0}^{(1)}(r)$ can be obtained as
$\psi_{0}^{(1)}(r)$=$[\psi_{0B1}^{(1)}(r),\psi_{0A1}^{(1)}(r),\psi_{0B2}%
^{(1)}(r),\psi_{0A2}^{(1)}(r)]^{T}$=$[f,g,ig^{\ast},-if^{\ast}]^{T}$ together
with Eqs. (\ref{transf}) and (\ref{solu}). From the relation $\mathcal{H}%
_{MF+}$=$-iI\mathtt{\otimes}\sigma_{2}\mathcal{H}_{MF-}iI\mathtt{\otimes
}\sigma_{2}$, another zero-mode solution is $\psi_{0}^{(2)}(r)$%
=$[ig,-if,f^{\ast},g^{\ast}]^{T}$. Note that the particular symmetry (defined
by the transformation $\Theta$) for $\mu$=$0$ ensures the $|n|$ zero modes,
unlike the single zero mode for odd vortex configuration for $\mu\neq$$0$
\cite{Gurarie}.

With the help of Eqs. (\ref{equs}), (\ref{equ2}) and (\ref{solu}), it is
interesting to see that the low energy behavior of our model is similar to
that of bilayer graphene. The analogy is not surprising, because after the
transformation $h_{MF}(\mathbf{k})$=$S^{+}\mathcal{H}_{MF}(\mathbf{k})S$ with
$S$=$\exp(i\pi I\mathtt{\otimes}s_{1}/4)\exp(i\pi I\mathtt{\otimes}%
s_{3}/4)\exp(i\pi I\mathtt{\otimes}s_{1}/2)$, our mean-field Hamiltonian has
the form
\begin{equation}
h_{MF}(\mathbf{k})=-\gamma_{0}(\gamma_{1}2\cos k_{x}+\gamma_{2}2\cos
k_{y}+me^{-i\chi\gamma_{5}}+\gamma_{0}\gamma_{5}\mu). \label{conh2}%
\end{equation}
Here in the Weyl representation, the Dirac matrices have the forms:
$\gamma_{\mu}$=$i\sigma_{2}\mathtt{\otimes}s_{\mu}$, $\gamma_{0}$=$\sigma
_{1}\mathtt{\otimes}I$, $\gamma_{\mu}$=$-i\gamma_{0}\gamma_{1}\gamma_{2}%
\gamma_{3}$=$\sigma_{3}\mathtt{\otimes}I$. The low-energy form of this
Hamiltonian at the two independent valleys is nearly identical to that for
bilayer graphene.

A localized zero mode in a gapped system with particle-hole symmetry usually
bounds a fractional charge $\pm e/2$ \cite{Hou,Franz,Jackiw,Chamon}%
\cite{wpsu,Goldstone}. Note that the Hamiltonian of our model in low-energy
limit is of the Dirac type with first-order differential operators. Therefore,
when $\mu$=$0$, and the OP has the form of a symmetric vortex of vorticity
$n$, i.e., Eq. (\ref{profop}), there are $\left\vert n\right\vert $
independent zero modes existing for each Dirac valley indicated by the index
theorem \cite{Weinberg}. When $\mu\mathtt{\neq}0$, the special symmetry
related to transformation $\Theta$ disappears and there is a single zero mode
(given by by Eqs. (\ref{solu}) and (\ref{transf})) for odd vorticity as argued
in \cite{Gurarie}. Other zero modes for $\left\vert n\right\vert \mathtt{>}1$
associated with one Dirac valley must split from zero energy and form in-gap
bound states which partially preserve zero modes properties just as pointed
out in Ref (9). Of course in general the inter-valley mixing would split the
pairs of zero modes. In this case, only axial charge fractionalization may
occur, which is in analogy with that in Ref. (13). The case of $\mu$=$0$ is an
exception due to the symmetry. The two zero modes belonging to different
valleys are degenerate and do not split in spite of the existence of
inter-valley mixing. Hence, each layer of the vortex core bounds a fractional
charge $\pm e/2$ when the two zero modes are occupied or not. Our numerical
calculation shown in the next section verifies the above picture.

\begin{figure}[ptb]
\begin{center}
\includegraphics[width=1\linewidth]{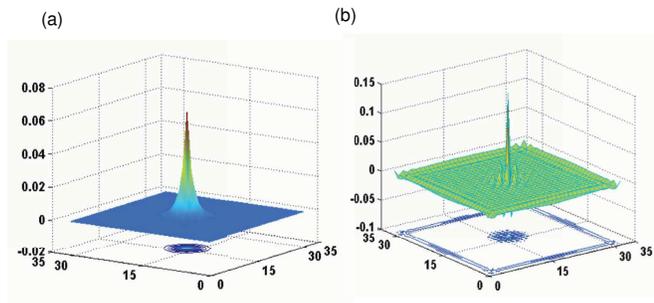}
\end{center}
\caption{(Color online) (a) Charge density distribution (with the uniform
background charge density distribution subtracted) for $n$=$1$, $\mu$=$0$ and
at half filling case with two zero modes occupied. The charge (the integral of
the charge density) is equal to $\frac{1}{2}e$. Here we give one layer case,
and the other layer has the same distribution. (b) The axial charge density
distribution with $n$=$1$ and $\mu$=$-$0.3. In all cases, we set the lattice
system size as $N$=$32\mathtt{\times}32\mathtt{\times}2$, $h$=$0$, and $U$%
=$4$.}%
\end{figure}

\begin{figure}[ptb]
\begin{center}
\includegraphics[width=1\linewidth]{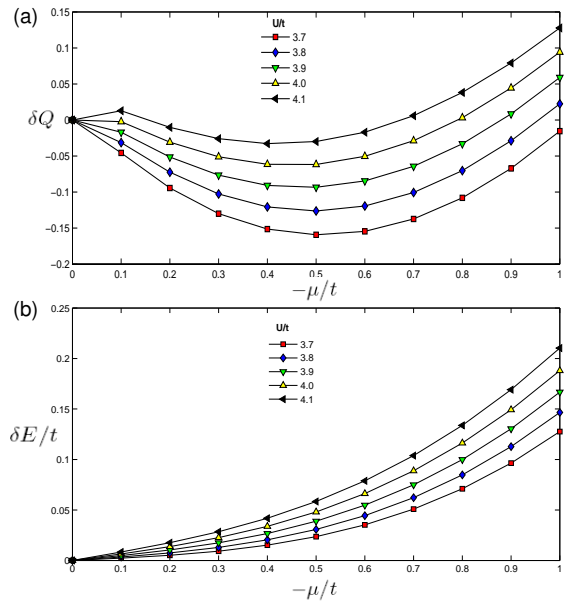}
\end{center}
\caption{(Color online) (a) The axial charge bound to a vortex profile as a
function of $U$ and $-\mu$. (b) Energy splitting [$E(N/2$+$1)-E(N/2)$] changes
as a function of $U$ and $-\mu$. In both cases, $h$=$0$ and the vortex profile
has vorticity $n$=$1$.}%
\end{figure}

\textbf{Numerics }We perform exact diagonalization of the mean-field
Hamiltonian of our system with $32\mathtt{\times}32$ sites per layer for
symmetric vortex profile of the OP in quantum limit with the amplitude of OP
calculated from the self-consistent solution of the BdG equations. In Fig. 2
we show the charge density distribution [Fig. 2(a)] and the axial charge [Fig.
2(b)] defined by $\delta Q_{\nu}$=$\sum_{i}[\left\langle \delta n_{i}%
\right\rangle _{\nu}\mathtt{-}\left\langle \delta n_{i}\right\rangle _{0}]$,
with subscript $\nu(0)$ referring to the system with $1(0)$ vortex and $\delta
n_{i}$=$n_{i2}\mathtt{-}n_{i1}$ the difference in the occupation number of the
two layers.

The numeric results show the \emph{rational } charge factionalization for
$\mu$\texttt{=}$0$ and \emph{irrational} fractional \emph{axial} charge for
$\mu\mathtt{\neq}0$. Here the rational fractional charge is exactly $\pm e/2$,
independent of $U$. The dependence of $\delta Q_{\nu}$ on $\mu$ and $U$ in
Fig. 3 (a) shows that the axial charge may be continuously tuned by changing
the external parameters.

The splitting of the zero energy levels due to inter-valley interaction is
shown in Fig. 3(b). The absence of splitting of zero energy levels for $\mu
$=$0$ can also be understood from a perturbative analysis. It can be seen that
the energy splitting has the form $\delta E\mathtt{\propto}|m_{0}\int
e^{-i(\mathbf{K}_{+}-\mathbf{K}_{-})\cdot\mathbf{{r}}}$Re$[e^{-i\theta}(f^{2}%
$+$g^{2})]d\mathbf{r}|$. One can see that $\delta E$=$0$ with the help of
solutions of Eqs. (14) and (16). Therefore, the inter-valley coupling does not
split the degenerate zero modes. The existence of the exact zero modes ensures
the charge fractionalization (per layer) as seen in Fig. 2 (a).

\textbf{Discussions }We propose a system of bilayer 2DEG adjacent to a type-II
superconductor thin film for EC. The vortex profile for OP can be generated by
defects or additional nontrivial texture in magnetic field \cite{jog}.
Furthermore, in our bilayer system, if a vacancy lies on a site in one layer
and meanwhile an interstitial induced by impurities occurs at the underlying
site in another layer, the nontrivial exciton OP with vortex profile is
generated due to its minimal coupling to the nonzero axial gauge field emerged
from the vacancy and interstitial relative to the magnetic flux lattices
background. Moveover, the axial fractional charge bounding a magnetic flux
quanta can form charge-flux composite particles which behave as Abelian anyons
with the exchange phase $\gamma$=$\pi\delta Q_{\nu}$ under the standard
argument \cite{Wick}. Note that the above studies have ignored the intralayer
next-nearest-neighbor hopping $t_{1}$ which may be important in the similar
systems \cite{Rosen,Hao}, where the fractional charge is related to
topologically nontrivial gapless edge states. However, the charge
factionalization discussed in the present paper is different from that case,
since the inter-layer interaction destroys the gapless edge states. Besides,
unlike the crude crystalloid, the influence of $t_{1}$\ here can be negligibly
small since the lattice distance of the Abrikosov vortex square array can be
adjusted very large.

The bilayer EC is analogous to the superconductor by the particle-hole
transformation. That is why the BdG equations were used. Furthermore, many
phenomena in superconductor have counterparts in bilayer EC, e.g., the vacancy
induced suppression of the amplitude of the OP \cite{salk}.

In summary, we have studied EC in a bilayer system which is likely to be
fabricated in laboratory. In the presence of topological non-trivial OP
profiles (vortex profiles), we have shown that the system exhibits
rational/irrational fractional (axial) charge.

\textbf{Acknowledgments} This work was supported by NSFC under Grants No.
90921003, No. 10874020, No. 10574150 and No. 60776063, by the National Basic
Research Program of China (973 Program) under Grants No. 2009CB929103, and by
a grant of the China Academy of Engineering and Physics.

\end{document}